# CoAP-DoS: An IoT Network Intrusion Data Set

Jared Mathews
*Cyber and Computer Sciences*
The Citadel
Charleston, USA
jmathew1@citadel.edu

Prosenjit Chatterjee
*Cyber and Computer Sciences*
The Citadel
Charleston, USA
chatterjee.prosenjit@citadel.edu

Shankar Banik
*Cyber and Computer Sciences*
The Citadel
Charleston, USA
baniks1@citadel.edu

*Abstract*—The need for secure Internet of Things (IoT) devices is growing as IoT devices are becoming more integrated into vital networks. Many systems rely on these devices to remain available and provide reliable service. Denial of service attacks against IoT devices are a real threat due to the fact these low power devices are very susceptible to denial-of-service attacks. Machine learning enabled network intrusion detection systems are effective at identifying new threats, but they require a large amount of data to work well. There are many network traffic data sets but very few that focus on IoT network traffic. Within the IoT network data sets there is a lack of CoAP denial of service data. We propose a novel data set covering this gap. We develop a new data set by collecting network traffic from real CoAP denial of service attacks and compare the data on multiple different machine learning classifiers. We show that the data set is effective on many classifiers.

## I. INTRODUCTION

IoT technology has been rapidly increasing in popularity as IoT has become the standard for most electronic devices in the home. IoT devices are lightweight appliances that usually have integrated sensors and communicate over the internet. They are used in smart homes, industrial control systems, medical systems, and much more. It is easy and cheap to implement IoT devices as they are typically compact and easy to produce. With a wide variety of devices and manufacturers, there does not currently exist any standards as IoT is an emerging technology.

As useful as IoT has become, there exist many vulnerabilities that can be introduced by implementing and relying on these lightweight devices. Many of the devices available are mass produced and lack common security measures used in more traditional computing devices. IoT devices are meant to be efficient and work in low power situations so often times security is overlooked in the manufacturing process. They lack the resources to prevent intrusion and attacks that more established technology can detect and stop.

One of the more devastating attacks IoT devices are vulnerable to is a denial of service (DoS) attack. This is where an attacker floods a device with traffic over a network in order to disable or overwhelm the device. IoT devices are particularly susceptible to DoS attacks due to the fact that it takes far less traffic to overwhelm them. In scenarios where IoT devices are used in industrial control or medical systems, a DoS attack can be devastating, far more so than in the case of smart home devices.

DoS attacks can usually be detected with network intrusion detection systems (NIDS) and prevented with good firewall rules. IoT devices are not capable of accommodating security apparatuses as they are only outfitted with enough power and resources to conduct their designed tasks. DoS attacks work by exploiting weaknesses in network communication protocols with the goal of flooding a device with too many packets for it to process along with its normal traffic load. The most common network protocols used by IoT devices are: a. Constrained Application Protocol (CoAP), b. Message Queuing Telemetry Transport (MQTT), c. Hypertext Transfer Protocol (HTTP), and d. Advanced Message Queuing Protocol (AMQP).

NIDS are used to detect DoS attacks over a network. NIDS can identify patterns in network traffic using predefined rule sets or artificial intelligence. Current advancements in the field of artificial intelligence, namely machine learning, have made it possible and effective to implement neural networks for intrusion detection. It is difficult to account for all possible attacks when defining an NIDS rule set, but with machine learning, unforeseen attacks can be identified that would have been overlooked in a rule set.

Large amounts of network data is needed for machine learning enabled NIDSs to identify DoS attacks against a node in a network. This network data can either be collected from the network to determine a baseline and any anomalies will be reported or sourced from public data sets with attack data. Public data sets can be used to train machine learning models to detect a wide range of network attacks including DoS attacks. There exist a wide array of public network traffic data sets for NIDS exist. KDD has a number of data sets for NIDS including KDD DDoS 2019 [1] and KDD DoS 2017 [2]. Both of these data sets are comprised of collections of multiple types of DoS attacks to be used with machine learning algorithms for NIDS. The KDD data sets lack data on DoS attacks against some of the common communication protocols used in IoT. The BotIoT data set [3], [4], [5], [6] focuses on network attacks against IoT devices. This data set focuses on MQTT traffic.

Currently there is not a large robust CoAP DoS data set for IoT NIDS. CoAP is a widely used protocol for IoT devices and very vulnerable to DoS attacks. In this research we propose a new robust CoAP DoS data set and evaluate it on many common machine learning algorithms. Realistic CoAP data is captured over 16 hours while two attackers conduct DoS attacks and a benign device sends normal traffic.

## II. RELATED WORK

Currently there are many publicly available network traffic data sets available with the collection of KDD data sets being the most well-known. KDD99 and NSL-KDD [7] are two of the more used KDD data sets. KDD99 is less relevant as NSL-KDD addresses issues with KDD99 and provides a more modern data set.

With regards to DoS attacks, the KDD collection contains the KDD DDoS 2019 [1] and KDD DoS 2017 [2]. KDD DDoS 2019 has a variety of DDoS attacks split into reflection and exploitation attacks. The data set has two day's worth of network traffic with 19 different DDoS attacks including attacks on both UDP and TCP. The data set was evaluated over multiple machine learning classifiers to determine the most useful features. KDD DoS 2017 focuses on application layer DoS attacks conducted over one day of experiments.

A NIDS data set is proposed in [8] to rival the NSL-KDD and KDD99 named UNSW-NB15. UNSW-NB15 was created using the IXIA tool to generate traffic. UNSW-NB15 is 100 gigabytes in size making it one of the largest data sets in the NIDS category. UNSW-NB15 is compared to NSL-KDD in [9], [10], [11]. UNSW-NB15 is shown to outperform NSLKDD making it less relevant in [9]. The data in UNSW-NB15 contains attacks against well-known protocols such as HTTP, file transfer protocol (FTP), and Secure Shell (SSH).

The previous data sets lack any attention on IoT data. The Bot-IoT data set [3], [4], [5], [5], on the other hand, focuses on network data from IoT networks utilizing the MQTT protocol. The Bot-IoT data set contains botnet attacks against a realistic IoT network infrastructure. Multiple types of attacks are provided including DDoS, DoS, and data exfiltration. [3] proposes a method of selecting features and adding new computed features. Bot-IoT contain 9543 benign packets and 73360900 attack packets.

Most of these data sets provide full pcap files of the network traffic. A data set is developed in [12] from publicly available NIDS data sets including UNSW-NB15, and Bot-IoT. The pcap data was processed into network flows and provides four new data sets created from four other existing data sets.

Another IoT data set is proposed in [13] for use with medical applications. A model for classifying malicious attacks on patient enabled IoT context aware devices used in a devices medical situation and a CoAP/MQTT data set were developed in [13]. IoT-Flock was used to generate the data set's traffic. Along with normal traffic, attack traffic was generated modeling the following attacks: MQTT Publish Flood, MQTT Authentication Bypass Attack, MQTT Packet Crafting Attack, and COAP Replay Attack. The data set was also evaluated over the following types of classifiers: Naive Bayes, K Nearest Neighbors, Random Forests, Matrix of Adaboost, Logistic Regression, and Decision Tree. While this data set does have CoAP data, it does not include CoAP DDoS attacks.

## III. PRELIMINARIES

### A. Recurrent Neural Networks

The process of emulating brain and neuron function in a computer with the goal of learning subtle patterns in data to make decisions is called deep learning. Deep learning techniques have recently become popular and have been successfully used in diverse domains. Neural networks use a system of nodes modeled similarly to neurons in the brain that utilize algorithms to learn patterns in data. A neural network generally consists of multiple layers of these nodes. The main layers are the input layer, hidden layers, and the output layer. The input layer is responsible for accepting the data that will be passed onto the subsequent layers. The hidden layers consist of algorithms used to find patterns in the data. The nodes in these layers use statistical analysis to find patterns in the data and assign weights to it. The final layer is the output layer. This layer is where the hidden layer passes the predicted outcome based on the weights generated. A RNN is a type of neural network that learns temporally or sequentially. The data passed into an RNN is sequentially and temporally related. The RNN architecture chosen is long short-term memory (LSTM). This model uses what is called a memory block. This mechanism takes the current input, the previous input, and memory of the previous input and decides whether or not to update the memory cell and output the state of the memory cell. This is useful in detecting DoS attacks because the time relationship between packets is the key factor in determining whether or not an attack is occurring.

TABLE I. EXPERIMENT MACHINE AND OS STRUCTURE

| Machine | OS | IP |
|---|---|---|
| Raspberry Pi | Raspbian GNU/Linux 9 (stretch) | 192.168.1.9 |
| Attacker PC 1 | macOS Catalina Version 10.15.7 | 192.168.1.12 |
| Attacker PC 2 | Ubuntu 20.04.1 LTS | 192.168.1.5 |
| Benign PC | Windows 10 Home Version 20H2 | 192.168.1.2 |

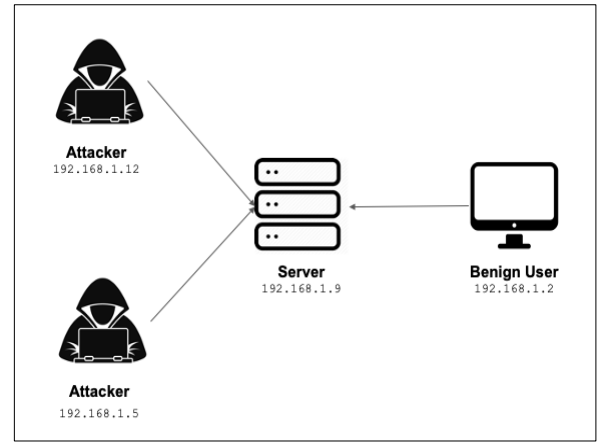

Fig 1. CoAP DDoS Network Structure

## IV. METHODS

### A. Data Set Creation

To create a useful CoAP DDoS data set, four devices were used. A raspberry pi as an IoT CoAP server, and three computer clients including two that were malicious. The server listened for any incoming packets which were tracked via Wireshark over 16 hours. The non-malicious client sent normal packets constantly to emulate normal traffic to an IoT device while the two malicious computers coordinated DDoS attacks at repeating intervals throughout the collection phase. The attack timestamps were recorded in json format for labeling the data for training. The final data set contains 661,304 benign and malicious packets which can be used to train a classifier. Figure 1 shows the network diagram of the experiment. The data set is publicly available at https://www.kaggle.com/jaredalanmathews/coapddos

The raspberry pi server was hosted at 192.168.1.9 in the experiment. The two attackers at 192.168.1.12 and 192.168.1.5 coordinate attacks against the server while the benign user at 192.168.1.2 sent normal traffic to the server which was hosted on port 8080. Table I contains the devices used including operating system and IP.

The benign client sent data of length 100 to 300 characters and would randomly alternate between GET, PUT, and POST requests. Between each delivery, the client would sleep for a random time between 2 to 7 seconds and then send another request, repeating until the predetermined time limit was over. The randomness simulates a more realistic situation for normal traffic.

TABLE II. ATTACK AMOUNTS AND TOTAL TIME

| Machine | Time (minutes) | Number Attacks |
|---|---|---|
| Attacker PC 1 Mac | 1000.68 | 410 |
| Attacker PC 2 Ubuntu | 1000.68 | 460 |

TABLE III. PACKETS FROM CLIENTS

| Source IP | Destination IP | Number Packets |
|---|---|---|
| 192.168.1.12 | 192.168.1.9 | 138011 |
| 192.168.1.5 | 192.168.1.9 | 123012 |
| 192.168.1.2 | 192.168.1.9 | 28964 |

The attacker clients could use two functionalities. The first mimics a benign client by randomly sending benign GET, PUT, and POST requests to the server. This attacker client also will randomly conduct a DoS attack which would rapidly send 300 PUT requests of data containing 9203 characters. The other attack functionality strictly conducts coordinated DDoS attacks that are timed to be sent at the same time as other attackers. Fig. II shows the number of attacks and time taken for both attackers.

The second attack functionality was used for this experiment. The two attacking clients began their DDoS attack every 10 minutes. Each attacking client produces a json file containing information about the attack for labeling purposes at the end of the timed attacks. Fig. III shows the breakdown of all packets from the clients. Attack packets make up for 39.47% of all of the packets in the pcap file where everything else is considered benign traffic.

---

**Algorithm 1:** Tokenization Algorithm

**Input:** 1D Matrix $a_m$ containing packet features, Matrix $B_{mn}$ with all non tokenized values from each packet

**Output:** Vector $\vec{b}$ with features tokenized

**for** $i = 0$ to $m$ **do**
    **if** $\vec{a_i} \notin B_i$ **then**
        $B_i \cup \vec{a_i}$
        $\vec{b_i}$n where $B_{in} = \vec{a_i}$
    **else**
        $\vec{b_i}$n where $B_{in} = \vec{a_i}$
    **end**
**end**

---

### B. Data Set Preprocessing

With the CoAP data set created, the data needed to be parsed to run through the neural network. The data was split into subsections determined by time. Each DoS attack conducted by the attackers consisted of 300 packets sent within 10 seconds and were coordinated at the same times.

The subsections of the data set were split into 10 second increments and were labeled malicious if they were in excess of 350 packets from the two malicious IP addresses.

The features that lend themselves to least bias needed to be determined after parsing the raw pcap data. To accomplish this, a genetic algorithm [14] was implemented to determine which features from a given .csv file are less likely to produce training bias. Fig. 2 shows the flow for the genetic algorithm used. Originally the packets were split into 42 features, and these were narrowed to 16 features after passing the data through the genetic algorithm. The features used were ethernet type, IP version, tos, length, id, flags, IP chksum, source port, seq, ack, dataofs, flags, window, UDP chksum, urgptr.

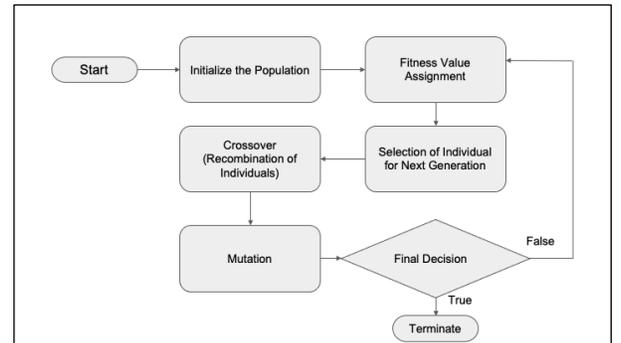

Fig 2. Genetic Algorithm Flow

**Algorithm 2:** Tensor Padding Algorithm

**Input:** Ragged Three Dimensional tensor: $T_{njm}$, max $j$ size: $n$, and packet for padding $p_m$
**Output** Padded Tensor $PT_{njm}$
**for** $i = 0$ to $n$ **do**
    **if** $T_{jm} < T_{nm}$ **then**
        **for** $n - j$ **do**
            $T_{jm} \cup \bigcup_{n=0}^{n-j} p_m$
        **end**
    **else**
**end**

**Algorithm 3:** Fronebius Normalization Algorithm

**Input:** 2D Matrix $A_{ij}$
**Output:** Normalized 2D Matrix $A'_{ij}$
$norm = [\sum_{ij} abs(A_{ij})^2]^{1/2}$
$A'_{ij} = A_{ij}/norm$

The packet features that were in string format needed to be tokenized. The arrays of packets that contained the flows were asymmetrical to each other and were padded to have uniform shape of all labeled data. Algorithm 1 shows the process for tokenizing this data and algorithm 2 shows the padding process.

The Frobenius norm was used to normalize the array of all packet values in each packet flow. This process can be seen in algorithm 3.

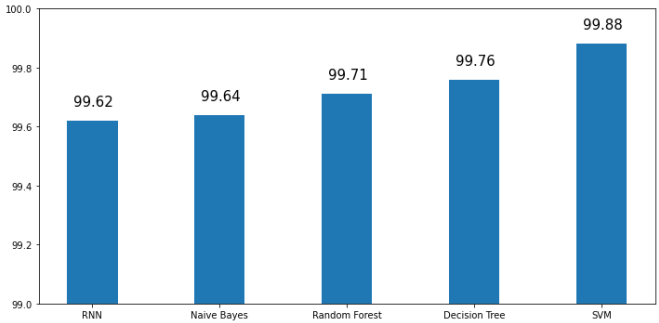

Fig 3.     Classifier Percent Success

### C. Data Set Evaluation

Five machine learning models were applied to the data set. The data set was trained on support vector machine (SVM), gaussian naive bayes, decision tree, random forests, and a recurrent neural network. Figs. 4, 5, 6, 7, and 8 show the confusion matrices of each of the algorithms on the CoAP data, and fig. 3 shows the percent successful predictions on test data. All of the classifiers were able to have a prediction success rate of over 99%. The SVM was overall the highest performing with 99.88% success. The RNN was surprisingly the lowest at 99.62% accuracy. Each of the classifiers perform exceptionally well on the CoAP data.

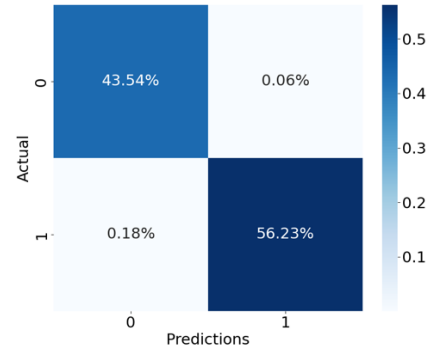

Fig 4.     Decision Tree Confusion Matrix

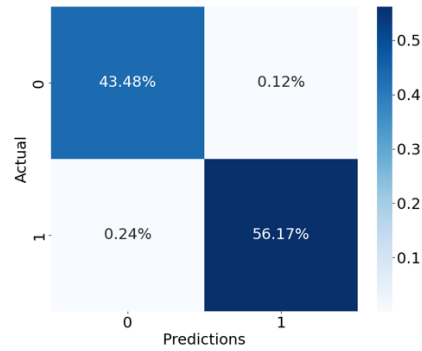

Fig. 5.     Naïve Bayes Confusion Matrix

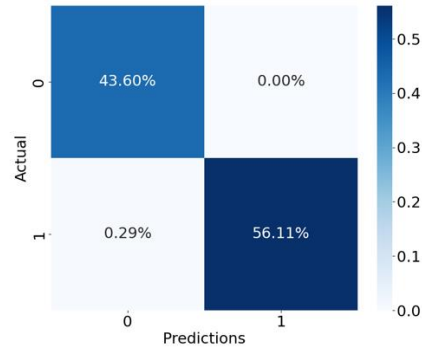

Fig. 6.     Random Forests Confusion Matrix

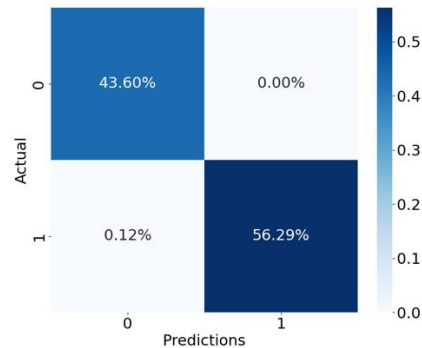

Fig. 7.     Support Vector Machine Confusion Matrix

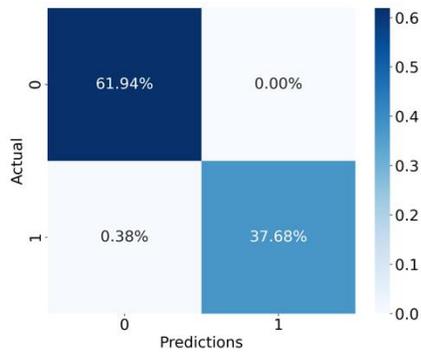

Fig. 8. Recurrent Neural Network Confusion Matrix

## V. Conclusion and Future Work

In this research we propose a new data set for use in classifying CoAP DoS attacks against IoT devices. While many other data sets exist with similar purposes, they fail to address specifically DoS attacks on the CoAP data set. This work fills that gap in hopes to increase IoT network security. Future work includes adding more attack types to the data set. While DoS attacks can be devastating, there do exist other attacks that pose real threats to IoT devices communicating over the CoAP protocol. More future work would be to evaluate the effectiveness of this data set in conjunction with the Bot-IoT data set. It would be interesting to see how well a classifier can detect CoAP vs MQTT attacks from training on both data sets.

## References

[1] I. Sharafaldin, A. H. Lashkari, S. Hakak, and A. A. Ghorbani, "Developing realistic distributed denial of service (ddos) attack dataset and taxonomy," in *2019 International Carnahan Conference on Security Technology (ICCST)*, 2019, pp. 1–8.

[2] "Detecting http-based application layer dos attacks on web servers in the presence of sampling," *Computer Networks*, vol. 121, pp. 25–36, 2017.

[3] N. Koroniotis, N. Moustafa, E. Sitnikova, and B. Turnbull, "Towards the development of realistic botnet dataset in the internet of things for network forensic analytics: Bot-iot dataset," 11 2018.

[4] N. Koroniotis, N. Moustafa, and E. Sitnikova, "A new network forensic framework based on deep learning for internet of things networks: A particle deep framework," *Future Generation Computer Systems*, vol. 110, pp. 91–106, 2020. [Online]. Available: https://www.sciencedirect.com/science/article/pii/S0167739X19325105

[5] N. Koroniotis and N. Moustafa, "Enhancing network forensics with particle swarm and deep learning: The particle deep framework," 2020.

[6] N. Koroniotis, N. Moustafa, F. Schiliro, P. Gauravaram, and H. Janicke, "A holistic review of cybersecurity and reliability perspectives in smart airports," *IEEE Access*, vol. 8, pp. 209802–209834, 2020.

[7] M. Tavallaee, E. Bagheri, W. Lu, and A. A. Ghorbani, "A detailed analysis of the kdd cup 99 data set," in *2009 IEEE Symposium on Computational Intelligence for Security and Defense Applications*, 2009, pp. 1–6.

[8] N. Moustafa and J. Slay, "Unsw-nb15: a comprehensive data set for network intrusion detection systems (unsw-nb15 network data set)," pp. 1–6, 2015.

[9] ——, "The evaluation of network anomaly detection systems: Statistical analysis of the unsw-nb15 data set and the comparison with the kdd99 data set," *Information Security Journal: A Global Perspective*, vol. 25, no. 1-3, pp. 18–31, 2016. [Online]. Available: https://doi.org/10.1080/19393555.2015.1125974

[10] N. Moustafa, J. Slay, and G. Creech, "Novel geometric area analysis technique for anomaly detection using trapezoidal area estimation on large-scale networks," *IEEE Transactions on Big Data*, vol. 5, no. 4, pp. 481–494, 2019.

[11] N. Moustafa, G. Creech, and J. Slay, Big Data Analytics for Intrusion Detection System: Statistical Decision-Making Using Finite Dirichlet Mixture Models. Cham: Springer International Publishing, 2017, pp. 127–156.

[12] M. Sarhan, S. Layeghy, N. Moustafa, and M. Portmann, "Netflow datasets for machine learning-based network intrusion detection systems," *Big Data Technologies and Applications*, p. 117–135, 2021.

[13] F. Hussain, S. Abbas, G. Shah, I. Pires, U. Fayyaz, F. Shahzad, N. Garcia, and E. Zdravevski, "A framework for malicious traffic detection in iot healthcare environment," *Sensors*, vol. 21, 04 2021.

[14] P. S. Muhuri, P. Chatterjee, X. Yuan, K. Roy, and A. Esterline, "Using a long short-term memory recurrent neural network (lstm-rnn) to classify network attacks," *Information*, vol. 11, no. 5, 2020. [Online]. Available: https://www.mdpi.com/2078-2489/11/5/243